\documentclass[12pt, a4paper]{article}
\usepackage{a4wide}
\usepackage[T1]{fontenc}
\usepackage{lmodern}
\usepackage{amssymb,amsmath}
\usepackage{ifxetex,ifluatex}
\usepackage{fixltx2e} 
\IfFileExists{upquote.sty}{\usepackage{upquote}}{}
\ifnum 0\ifxetex 1\fi\ifluatex 1\fi=0 
  \usepackage[utf8]{inputenc}
\else 
  \ifxetex
    \usepackage{mathspec}
    \usepackage{xltxtra,xunicode}
  \else
    \usepackage{fontspec}
  \fi
  \defaultfontfeatures{Mapping=tex-text,Scale=MatchLowercase}
  
\fi
\IfFileExists{microtype.sty}{\usepackage{microtype}}{}
\usepackage[margin=1in]{geometry}
\usepackage{graphicx}
\makeatletter
\def\ScaleIfNeeded{%
  \ifdim\Gin@nat@width>\linewidth
    \linewidth
  \else
    \Gin@nat@width
  \fi
}
\makeatother
\let\Oldincludegraphics\includegraphics
{%
 \catcode`\@=11\relax%
 \gdef\includegraphics{\@ifnextchar[{\Oldincludegraphics}{\Oldincludegraphics[width=\ScaleIfNeeded]}}%
}%
\ifxetex
  \usepackage[setpagesize=false, 
              unicode=false, 
              xetex]{hyperref}
\else
  \usepackage[unicode=true]{hyperref}
\fi
\hypersetup{breaklinks=true,
            bookmarks=true,
            pdfauthor={Prof.dr. Richard D. Gill (Leiden University)},
            pdftitle={Rarity of Respiratory Arrest in ED?},
            colorlinks=true,
            citecolor=blue,
            urlcolor=blue,
            linkcolor=magenta,
            pdfborder={0 0 0}}
\begin{document}

\title{Rarity of Respiratory Arrest in ED?}
\author{Richard D. Gill\\ Mathematical Institute\\ Leiden University\\
\small \url{http://www.math.leidenuniv.nl/~gill}}

\date{9 July, 2014}

\maketitle
\begin{abstract}
\noindent Statistical analysis of monthly rates of events in around 20 hospitals
and over a period of about 10 years shows that respiratory arrest, though
about five times less frequent than cardio-respiratory arrest, is a common
occurrence in the Emergency Department of a typical smaller UK hospital.
\end{abstract}
\tableofcontents

\section{Summary}\label{summary}

In the Ben Geen case, three experts (an anaesthetist, a charge nurse,
and a head nurse at three different hospitals) have given their opinion
that primary respiratory arrest in ED (Emergency Department) is rare.
The defence, following the eminent statistician Prof.~Jane Hutton, argue
that these statements at best merely constitute anecdotal evidence and
at worst can be strongly tainted by observer bias. F.O.I.\ requests have
consequently been made to many hospitals similar to Horton General,
resulting in a large data-base containing numbers of various events in
ED as well as total numbers of patients admitted to ED per month, in
more than 20 hospitals and covering up to 10 years. This report
describes the main findings from statistical analysis of the data-base.

We find that respiratory arrests in ED are about five times less
frequent than cardio-respiratory arrests, which are of course extremely
frequent. Respiratory arrest is certainly less common than
cardio-respiratory, but certainly not rare at all, by any reasonable
understanding of the meaning of the word ``rare''.

The \emph{relative} size of the variation in small observed rates due
purely to chance is much, much larger than the \emph{relative} size of
the variation in larger observed rates (The ``law of small numbers'' --
Poisson variation). On top of purely random variation and strong
seasonal variation, the numbers fluctuate quite wildly in time,
exhibiting all kinds of trends, bumps, and gaps in different hospitals.

Altogether, one can only conclude that periods with ``surprisingly high
numbers'' of respiratory arrests are by no means rare and hence not in
themselves surprising at all. The quality of the data (which to put it
kindly, is not high) moreover underlines that all kinds of
classification and reporting issues could easily go some way to explain
these fluctuations. How events are classified, and how patients are
admitted, will vary in time as hospital policies change; moreover,
random fluctuations in numbers of events can trigger changes in how
events are classified (so called ``publicity bias'').

To sum up: respiratory arrest in ED is not rare at all, and moreover its
frequency is subject to large, and to a large extent unpredictable,
variation of quite innocent nature.

The data-base analysed in this report is available at
\url{http://www.math.leidenuniv.nl/~gill/Data/Hdf.csv}, 
and the statistical analysis scripts 
(written in the \texttt{R} language for statistical computing) can be
inspected at \url{http://rpubs.com/gill1109/DraftOpinion}.
A table with an overview of hospitals and trusts to which F.O.I.\ requests 
were submitted is reproduced in the appendix.

\section{Preparation of the data}\label{preparation-of-the-data}

This report focusses on a subset of 16 hospitals (or hospital trusts).

Originally, F.O.I.\ requests were sent to around 30 different hospitals
and/or hospital trusts, supposed to be similar to Horton 
(though a few are teaching hospitals about five times larger). 
A few hospitals did not respond or turned out no longer to exist. The
hospitals and trusts which did submit any data used a multitude of
different formats including pdf files which though in one sense digital,
are actually totally unsuited for transferring large tables of numbers
from a hospital data base to a statistician's computer.

After an extremely laborious process we succeeded in building a more or
less ``clean'' data-base \url{http://www.math.leidenuniv.nl/~gill/Data/Hdf.csv}
in a format amenable to statistical analysis
corresponding to 22 hospitals or trusts. This means that data from eight
trusts did not make it into the present data-base for various
administrative reasons, the most common reason being
that the data asked for was simply not available. The second most common
reason was accidental error on our side (lost emails!). We will be able to
rectify some of these omissions later, which will add a small number of 
hospitals to the data-base, but we do not believe this will have any impact 
on our main conclusions. 

Data from \textbf{Horton General Hospital is not included}: all these
analyses have been performed before even obtaining any data from that
hospital.

For the initial analyses in this report, six hospitals from the
data-base have been removed: the data on one of those hospitals is quite
weird, the other five did not supply the monthly number of admissions to
ED. In an appendix we show what happens when those five are put back:
nothing much changes.

\section{What are we measuring}\label{what-are-we-measuring}

In this report, we will study three variables called here
\texttt{Admissions}, \texttt{CardioED} and \texttt{RespED}. The original F.O.I.\ 
requests defined these data as follows:

\texttt{Admissions}: The number of patients admitted to
\emph{Hospital/Trust X} Emergency Departments, by month, from November
1999 to the present.

\texttt{CardioED}: The number of patients admitted to
\emph{Hospital/Trust X} Critical Care Units with cardio-respiratory
arrest from the Emergency Department, by month, from November 1999 to
the present.

\texttt{RespED}: The number of patients admitted to \emph{Hospital/Trust
X} Critical Care Units with respiratory arrest from the Emergency
Department, by month, from November 1999 to the present.

The variable \texttt{Admissions} therefore counts total admissions \emph{to}
ED, and gives us information about the size of the hospital. Moreover,
we are specifically interested in events happening \emph{in} ED which
lead to transfer to CC (Critical Care units, including Intensive Care
units). Therefore ``size of ED'', as measured by rates of admission to
ED is more relevant than ``size of hospital'' measured in number of
beds, say.

For completeness, I mention that \texttt{CardioEd} and \texttt{RespED} are just
two of a collection of altogether six variables whose names are formed
by combining a prefix \texttt{Cardio}, \texttt{Resp}, or \texttt{Hypo} with a suffix
\texttt{ED} or \texttt{Tot}. The suffix \texttt{ED} stands for Emergency Department
(Accident and Emergency, A\&E): the number of such admissions which are
\emph{from} ED. The suffix \texttt{Tot} stands for total: the total number of
admissions to critical care units from anywhere, with the corresponding
diagnosis.

The prefixes \texttt{Cardio}, \texttt{Resp}, and \texttt{Hypo} stand for cardiac or more
precisely, cardio-respiratory arrest, respiratory arrest, and
hypoglycaemic arrest.

The intention was that the variables \texttt{CardioED}, \texttt{RespED}, and \texttt{HypoED}  
should count \emph{events in ED causing transfer to CC},
rather than \emph{diagnosis (events in the recent medical history) of
the patient when transferred from ED to CC}. In other words, they were
intended to count events occurring \emph{after} the patient was admitted
to ED, whose occurrence \emph{in} ED was the direct \emph{cause} of
transfer from ED to CC. This should be compared to the variables
 \texttt{CardioTot}, \texttt{RespTot}, and \texttt{HypoTot}, which were intended to count
patients entering CC with the respective diagnoses, irrespective of when
the corresponding event had occurred and what was the immediate reason
for the admission to CC.

We can only hope that most hospitals did interpret the F.O.I.\ request as
intended. A number of hospitals did not supply any information on the
numbers of events occurring in ED -- they could only supply data at
higher or different aggregation levels. This means that our key variables
\texttt{CardioED} and \texttt{RespED} were often missing. 
For similar reasons, the variable \texttt{Admissions} also was often missing.
It was sometimes not easy
to see from submitted spread-sheets and supporting documentation whether
a blank stood for ``zero'' or ``not available''. 

It is not entirely clear whether any particular patient only has one
diagnosis, or can have several. Cardio-respiratory arrest is
heart-failure (cardiac arrest) together with respiratory failure because
the former caused the latter. When your heart stops beating your lungs
rapidly stop breathing, so a cardiac arrest without respiratory arrest
is essentially impossible, except perhaps in IC (think of a patient in a
breathing machine).

Suppose a patient comes into ED who has already been resuscitated after
a cardiac arrest. Suppose this patient subsequently (while in ED) also
suffers a respiratory arrest. He or she now has both \emph{diagnoses}
(both these things have recently happened to him or her). If this
patient is now admitted to CC, is he or she counted both as an
admittance to CC with cardio-respiratory arrest and as an admittance to
CC with respiratory arrest?

The intention was that cardio-respiratory and respiratory arrest should
be mutually exclusive categories, but the F.O.I.\ does not make that
explicit, though one can consider it implicitly implied when one
considers all seven questions together. Fortunately, we will be able to
take finesse this particular difficulty.

What is respiratory arrest and what is cardiac arrest, anyway? 
An expert tells me \begin{quote}
\emph{Both are merely symptoms of an underlying problem. 
For example a mid-brain stroke may result in respiratory arrest, 
which leads on to cardio-respiratory arrest if not treated – the heart stopping 
if artificial respiration has not been instituted.
So if the medics pick up early on the stroke, it may only get as far as a fall. 
If the stroke was left untreated respiratory arrest may follow. 
If that is left untreated, the heart stops also. 
So the diagnosis is stroke, and the outcome arrest.}
\end{quote}

Wikipedia redirects from cardio-respiratory arrest to cardiac arrest.

\section{Summary of issues around
diagnosis}\label{summary-of-issues-around-diagnosis}

In the previous section I have discussed difficulties interpreting the
data revolving around the fact that the condition(s) a patient has when
transferred to CC is not the same as the immediate cause of the
transfer. In principle, a patient can have experienced both
cardio-respiratory and respiratory arrest, in either order. These events
can happen before admission to the hospital Emergency department, or
during stay in Emergency. Possibly, one event led to admission to ED,
the next event to transfer to CC. What we wanted to count were transfers
to CC caused directly either by a respiratory arrest in ED, or by a
cardio-respiratory arrest in ED.

This all turns around the difference between the little words ``in'' and
``with'', and whether, when one asks for numbers of patients in
different categories, administrators (or their data-base software) will
understand that the categories should be understood as mutually
exclusive. It depends on what information actually is in the data-base.
I do not know how the the F.O.I.\ requests have been interpreted by the
hospital administrators who have kindly supplied us all this data. We
can go back and ask. Or we can ask medical experts what they think those
questions actually mean, and what data they think these questions would
actually elicit.

On the other hand, we are missing cardio-respiratory and respiratory
arrests in ED which do not result in transfer to CC, if that is
possible. Many events occur in hospital wards which do not end up in the
hospital data-base. If a patient has suffered either arrest in ED and is
immediately successfully resuscitated there, does that person necessarily
go directly to a critical care unit?

Finally we should be aware that the records stored in hospital
data-bases were not collected for the purpose of answering our
questions, but are the results of an administrative system which
collects some information about some of the processes going on in the
hospital, but not all. Many events do not find their way into the
data-base at all. Many events are wrongly classified. In any case, the
classification can be somewhat subjective. The system allows only a
small collection of possible categories and choosing just one of them
might well not do justice to the complex state of any particular
patient. So an administrator picks one out of habit or for convenience.
Registered rates of various kinds of event can change because culture
changes, policy changes, staff changes, staff start ``seeing'' a new
kind of event happen more often because they have been alerted to it by
a notable occurrence; thus awareness of particular categories of events
changes in response to occurrence of other events.

\section{Which Hospitals?}\label{which-hospitals}

``Hospital'' is the name (more precisely: \emph{my} ``short name'') of
the hospital, or in some cases the trust. A table will be supplied
separately, giving full names of hospitals and trusts.

Here are the (short) names of the 22 hospitals (or trusts) in our data-base: 
\begin{verbatim}
Barking          C Manchester    Darlington       Doncaster & B    
Frenchay         Good Hope       Heartlands       Hertfordshire    
Hexham           Hull & E Yorksh Leicester        Maidstone       
N Tyneside       Nottingham      Oxford Radcliffe R Liverpool      
Sandwell         Solihull        Stoke            UHN Durham       
Wansbeck         Wycombe
\end{verbatim}
These are the names of
the 16 hospitals left after we have omitted those which did not report
the numbers of admissions to ED:
\begin{verbatim}
C Manchester     Doncaster & B   Frenchay         Good Hope        
Heartlands       Hexham          Hull & E Yorksh  Leicester        
Maidstone        N Tyneside      Nottingham       Oxford Radcliffe
R Liverpool      Sandwell        Solihull         Wansbeck
\end{verbatim}
The six hospitals which have been omitted to form the smaller data set
are 
\begin{verbatim} 
Barking          Darlington       Stoke           UHN Durham          
Wycombe          Hertfordshire
\end{verbatim}
The first five, because the number of admissions to ED is
missing; the sixth, Hertfordshire, because the numbers there do not make
much sense at all, and probably had not been processed correctly.

As mentioned before, it was actually very hard to deduce whether blank
fields in tables of numbers in the files provided by some hospitals
meant ``zero'' or ``not available''. As we will later see, another four
hospitals (Sandwell, Solihull, Heartlands, Good Hope) need to be removed
from the presently remaining 16 for this reason. On the other hand, for
the final steps of our analysis below, we will not use ``total admissions to ED''
so we could just as well have put Barking, Darlington, Stoke, UHN
Durham and Wycombe back in. We will do that in an
appendix. \emph{It turns out that our substantive conclusions do not
change at all}.

\section{Which Time Period?}\label{which-time-period}

We have \emph{monthly data} from each hospital from various periods of
time, but all within the overall period November 1999 to December 2011.
That is 12 years and 2 months, or altogether 146 months. The variable
\texttt{MonthNr} in our analyses measures time, by months, starting with
month -1 = November 1999, month 0 = December 1999, month 1 = January
2000, \ldots{}, month 144 = December 2011.

Most hospitals could only supply data for (varying) parts of the period
named in the F.O.I.\ request. This will be clearly visible in the graphics
shown later in this report.

\section{Are the hospitals
comparable?}\label{are-the-hospitals-comparable}

I have deliberately avoided studying data from Horton General Hospital, 
in order to avoid personal bias. Roughly speaking, the hospitals
in this study vary in size by a factor of up to 5: we have quite a few
hospitals with around 250 beds, quite a few around 500, a few with around
750, and just a couple with more than 1000 beds. Horton General belongs
at the low end of the scale, among Hexham, Solihull, Wansbeck, Wycombe.

\section{Visualisation of the Data}\label{visualisation-of-the-data}

For the time being we look at 16 hospitals, partly for the opportunistic
reason that $16 = 4 \times 4$, which is very convenient for graphical displays
in which we can see the individual data of each hospital separately.

We plot just three of our variables against time (\texttt{MonthNr}). The
three variables of interest in this report are \texttt{Admissions},
\texttt{CardioED}, and \texttt{RespED}.

\bigskip
 
\noindent
\includegraphics{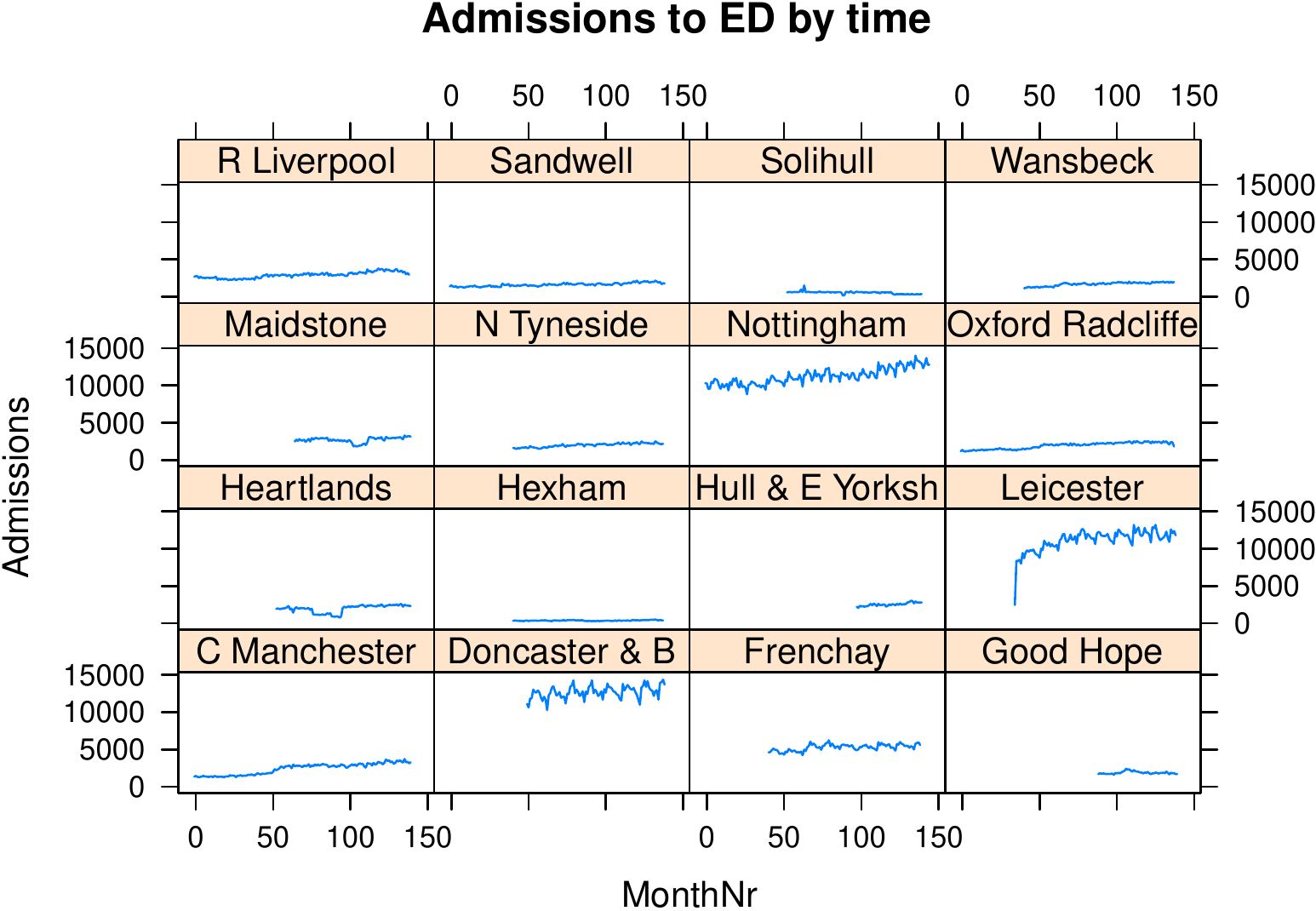}

\bigskip

Three hospitals -- Nottingham, Leicester, Doncaster \& B -- stand out as
having apparently five times larger emergency departments than most of
the others: the three big ones have around 11\,000 admissions per month
(mean monthly admissions equals 11\,196);
the smaller ones only around 2\,000 (mean monthly admissions of all remaining
hospitals equals 2838, or around 3\,000). The regular seasonal fluctuations in
the large admission numbers are particularly clear.

Nottingham and Leicester are both big teaching hospitals. Doncaster \& B
is a trust: my short name is short for ``Doncaster and Bassetlaw
\emph{hospitals}''.

I draw the plot again, capping the admissions at 6000, so we can better
see the 13 time series of lower numbers. Seasonal variation at Frenchay
is very clear to see; less visible in the others. This too is only to be
expected: by the law of \emph{small} numbers (Poisson variation if not
super-Poisson variation) random variation becomes proportionately larger
when looking at low rates, hence more easily masks a given amount of
systematic variation.

\bigskip

\noindent 
\includegraphics{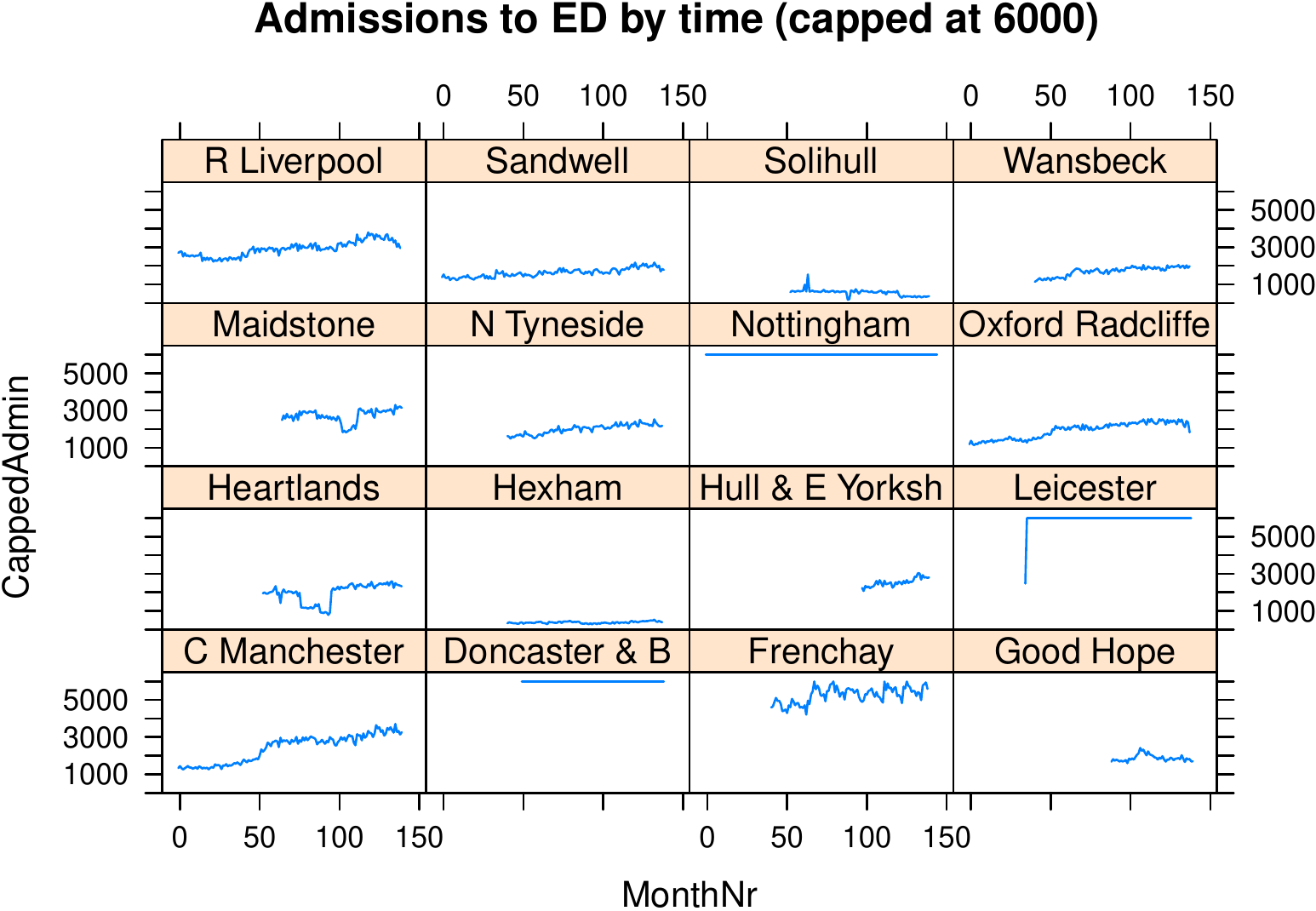}

\bigskip

Now we turn to the heart of the matter: admissions to CC (Critical Care
units, Intensive Care) from ED (Emergency Department, \emph{aka} A\&E)
because of (or at least: with) cardiac and/or respiratory arrest.

The \emph{intention} was to count admissions to CC from ED caused by
just one of those events. If both had occurred, then the first might be
reasonably imagined to have triggered the second. In other words, we
wanted to know the numbers of admissions to CC caused \emph{primarily}
by either type of arrest having occurred \emph{after} admission to ED.
However we do not know how exactly the hospitals have interpreted the
request for data, or indeed, whether the interpretation was uniform.
Fortunately, whether the counts are of cases ``with'', or only cases
``primarily caused by'', we will still be able to extract some very
pertinent information from the data.

\bigskip

\noindent
\includegraphics{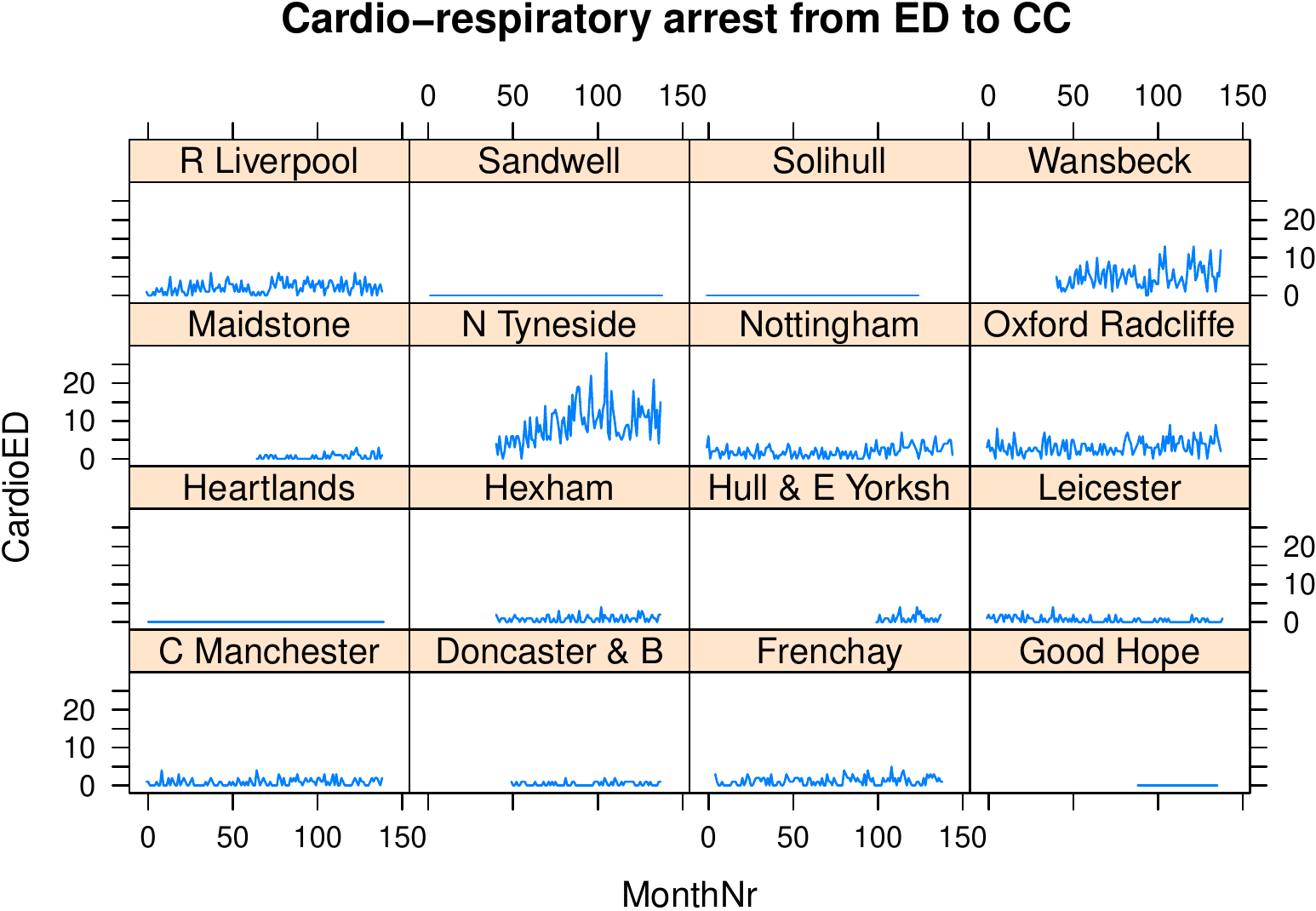}

\bigskip
\bigskip
\bigskip

\noindent
\includegraphics{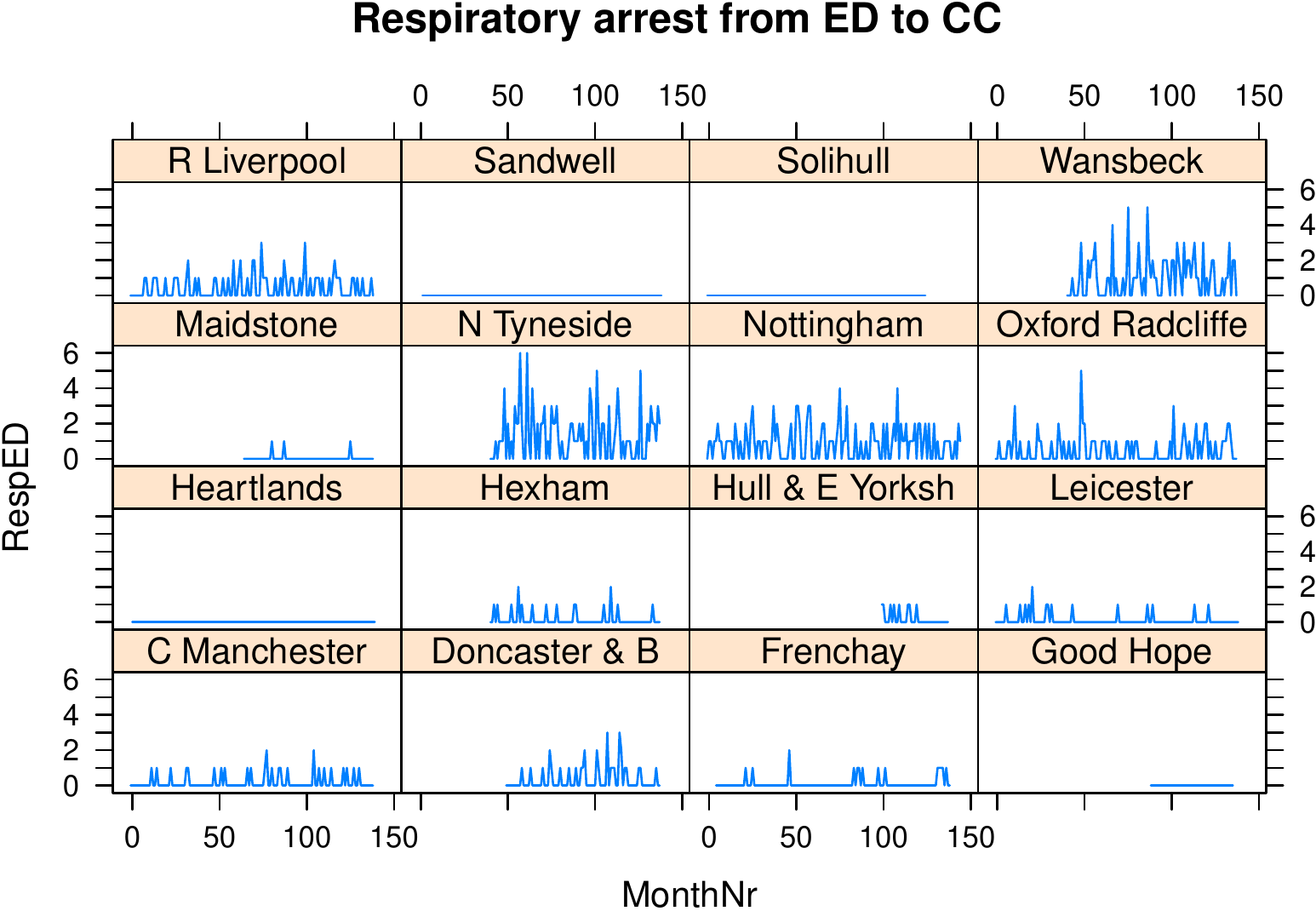}

\bigskip

\section{Observations}\label{observations}

Very globally, we can say that even in the smaller hospitals there often
1 or 2 respiratory arrests in one month (sometimes none, sometimes 3 or
4), and anything from 0 to 10 and upwards cardio-respiratory arrests.
Transfer from ED to CC because of cardio-respiratory arrest is, on the
whole, very common. Transfer from ED to CC because of respiratory arrest
is less common but not rare, by any account.

Four important features should be observed.
\bigskip

\noindent
\textbf{First feature}: four hospitals stand out as not reporting any
cardiac or respiratory arrests at all in ED: Sandwell, Solihull,
Heartlands, and Good Hope.
\bigskip

\noindent
\textbf{Second feature}: cardio-respiratory arrest is about five times
more frequent than respiratory arrest. It is therefore somewhat less
common. But it could well be considered rather misleading to call it
``rare''.
\bigskip

\noindent
\textbf{Third feature}: within each hospital, the numbers per month are
highly variable.
\bigskip

\noindent
\textbf{Fourth feature}: there are clear differences in levels between
different hospitals, up to perhaps a factor of 5 between the lowest and
the largest numbers. Regarding the numbers of admissions to ED this is
mainly accounted for by scale. Regarding the numbers of transfers to CC
because of (or with) various diagnoses this is no doubt exacerbated by
different interpretations of the events to be counted, different
registration systems or cultures. This careful selection of ``similar''
hospitals is actually extremely inhomogenous, even taking account of
scale (size). Inhomogeneity might be administrative and/or cultural in
nature, rather than due to scale or case-mix differences.

\bigskip

Obviously, we should compare Horton General to \emph{similar} hospitals.
Regarding size, this means Hexham, Solihull, Wansbeck, and Wycombe. As
we will see the data from Solihull is anomalous, so this leaves us with
Hexham, Wansbeck, and Wycombe.

\section{Dealing with Anomalies}\label{dealing-with-anomalies}

The complete absence of cardiac arrest in Sandwell, Solihull,
Heartlands, and Good Hope \emph{must} be caused by data-registration
issues. It is inconceivable that there was not a single
cardio-respiratory arrest in ED in all those years. I suspect we have
incorrectly interpreted ``blank'' columns in a spreadsheet as zeroes
rather than ``not available''.

So in the next section, I will remove the hospitals with \textbf{zero}
events -- I am guessing that these are not true zeroes, but rather ``not
known''. In any case, hospital months when \emph{neither} event happens
do not tell us anything about whether respiratory without cardiac arrest
is rare.

\section{Some Statistics}\label{some-statistics}
The following statistics therefore pertain to just 12 hospitals
\begin{verbatim}
C Manchester     Doncaster & B    Frenchay         Hexham           
Hull & E Yorksh  Leicester        Maidstone        N Tyneside
Nottingham       Oxford Radcliffe R Liverpool      Wansbeck
\end{verbatim}
\bigskip
\noindent Mean number of respiratory arrests per month:
\begin{verbatim}
0.4592
\end{verbatim}

\bigskip
\noindent Mean number of cardio-respiratory arrests per month:
\begin{verbatim}
2.207
\end{verbatim}

\bigskip
\noindent Number of hospital months in which the number of respiratory arrests exceeded the number of cardio-respiratory arrests:

94

\bigskip
\noindent Total number of hospital-months in the data from this sample of 12 hospitals:

415

\bigskip
\noindent Average number of months per year in which the number of respiratory arrests exceeded the number of cardio-respiratory arrests:

2.718

\bigskip
\noindent Average number of respiratory arrests per year:

5.511

\medskip
\noindent
\textbf{In round numbers, cardio-respiratory arrest is five times more common that respiratory arrest.}

\medskip
\noindent
\textbf{\emph{Even if} some patients are counted twice, \emph{at least} half of the respiratory arrests 
are without accompanying cardio-respiratory arrest.}

\section{Analysis}\label{analysis}

In very round numbers, per hospital, there are on average about \emph{3
months} in every year with a respiratory but no cardiac arrest.
Therefore there are  \emph{at least 3 cases} per year of respiratory without
cardiac arrest. There are on average about 6 respiratory arrests per
year. This means that \textbf{\emph{at least half} (if not all) of the respiratory arrests
are \emph{not} accompanied by cardiac arrest}.

On average, per hospital, there are \textbf{at least} about 3
respiratory arrests (unaccompanied by cardiac arrest) per year; that can hardly be called \emph{rare}.
It is true, but hardly relevant, that \emph{respiratory arrest is less
common than cardiac arrest} (about five times as infrequent).

\section{Conclusion}\label{conclusion}

Though some of what is called ``respiratory arrest'' in our data-sets
might actually represent a combination of respiratory and cardiac
arrest, in either order, it is absolutely clear that \textbf{respiratory
arrest (not caused by immediately preceding cardiac arrest)} (and
leading to transfer to CC) is not \emph{rare} at all.

Respiratory arrest -- leading to transfer to CC -- is about five times
less frequent than cardio-respiratory arrest. In a hospital of the same
size as Horton General, there are on average 1 or 2 cases per month.
Fluctuations are large.

Respiratory arrest \emph{not} leading to transfer to CC has not been
accounted for at all: the numbers are unknown, unregistered.
\small
\section{Appendix 1: Changing the sample of hospitals}\label{appendix}

Let's check what happens when we put back the hospitals with no
``admissions to ED'' data. That means we are now talking about the
following 17 hospitals; and our sample now has relatively more smaller
hospitals.
\begin{verbatim}
Darlington       Frenchay         Good Hope        Hertfordshire    
Hexham           Hull & E Yorksh  Leicester        Maidstone        
N Tyneside       Nottingham       Oxford Radcliffe R Liverpool     
Sandwell         Solihull         Stoke            UHN Durham       
Wycombe
\end{verbatim}

\bigskip
\noindent Mean number of respiratory arrests per month:

0.3077

\bigskip
\noindent Mean number of cardio-respiratory arrests per month:

1.563

\bigskip
\noindent Number of hospital months in which the number of respiratory arrests exceeded the number of cardio-respiratory arrests:

84

\bigskip
\noindent Total number of hospital-months in the data from this sample of 12 hospitals:

854

\bigskip
\noindent Average number of months per year in which the number of respiratory arrests exceeded the number of cardio-respiratory arrests:

1.18

\bigskip
\noindent Average number of respiratory arrests per year:

3.693

\medskip
\noindent
\textbf{In round numbers, cardio-respiratory arrest is still five times more common that respiratory arrest.}

\medskip
\noindent
\textbf{Even if some patients are counted twice, at least one third of the respiratory arrests 
are without accompanying cardiac arrest.}

\bigskip

The hospitals we put back, all of them quite small, 
have reduced the overall rates both of cardiac
and of respiratory arrest. However, we still see that with respect to
this larger sample of hospitals, including more small hospitals,
respiratory arrest without cardiac arrest accounts for \textbf{at least}
one third of (if not all) respiratory arrests; and respiratory arrest, though less
common that cardio-respiratory arrest (it occurs five times less
frequently), still occurs many times year. It cannot be called \emph{rare}.

\section{Appendix 2: The author's expertise}

I am a mathematician and a statistician, presently full professor of mathematical statistics at Leiden University, Netherlands (Mathematical Institute, Science Faculty). I am presently 62 years old. I have both British and Dutch nationality. I am a member of the Royal Dutch Academy of Sciences, and a past president of the Dutch Statistical Society, to mention just two marks of distinction. My research interests span both theoretical and applied statistics. I have worked for a long time in medical statistics, both on topics connected to clinical trials and to observational studies (epidemiology). This work has involved many collaborations with (hospital) medical doctors.

More recently I became involved in forensic statistics which is the art and science of applying statistics and probability to problems of two kinds: statistics involved in solving crimes (police investigation) and statistics involved in prosecuting criminals (evaluating the weight of statistical evidence). For instance, I have recently worked for the United Nations Special Tribunal on Lebanon, analysing mobile phone meta-data used to identify (?) the perpetrators of the assassination of Prime Minister Hariri some years ago. I am now regularly consulted by the Dutch police and by Dutch courts. Recently I was asked by a Dutch court to collaborate with a gynaecologist in order to comment on probabilities in a case of (alleged) serial infanticide. It was absolutely necessary for a medical expert and a statistical expert to look at the evidence and the relevant scientific literature \emph{together}. We needed to figure out \emph{what were the right questions to ask}, and neither of us could do that on our own. Lawyers and judges are even worse placed to figure out \emph{what are the right questions to ask}. Fortunately, the realisation that multi-disciplinary scientific work should be performed in first instance by collaborating scientists, not by courts of law, is growing, due to many recognised miscarriages of justice where faulty interpretation of scientific evidence, and recruitment of the ``wrong'' scientific experts, has been involved.

Particularly relevant to the present particular case (Ben Geen) is my involvement in a celebrated Dutch miscarriage of justice. A nurse, Lucia de Berk, was given a life sentence for murder of 6 patients and attempted murder of 4 more, largely on the basis of statistical evidence linking her presence to ``incidents'' on the wards where she worked. The conviction was revoked after a sequence of legal battles lasting altogether 9 years. At the final acquittal, the judges not only announced that she was not guilty, but that in actual fact no murders had occurred at all. In actual fact, according to the trial judges, nurses had battled heroically to save lives of patients which, despite this, were ultimately shortened by the \emph{mistakes of their doctors}. Cases like this, internationally, are by no means rare. In fact, ``health care serial killers'' are rare, but witch hunts triggered by medical errors and magnified out of proportion by the rigid social structure of a hospital are unfortunately all too common, with often devastating consequences.

\section{Appendix 3: Some anecdotes from the case of Lucia de Berk}

The case of Lucia de Berk -- perhaps the biggest miscarriages of justice which ever occurred in the Netherlands, a country which prides itself on its justice system -- contains a multitude of shocking parallels with the case of Ben Geen. What is all the more shocking is that this seems to have gone totally unremarked, to date.

I will here recall just a few ``anecdotes'' from that case which have particular relevance to the statistical aspects of the Ben Geen case. 

A key piece of evidence in the Lucia case was that the number of incidents on Lucia's ward was 9 in one year  (the year in which she was supposed to be on a killing spree), and close to zero during both the two preceding years, and in the subsequent year. This enormous unexpected number of incidents in that ward was a key piece of prosecution evidence. It later transpired that the name of the ward had been changed, just prior to those two years of almost no incidents. In the years preceding, it had been somewhat larger than 9, several years in succession. The hospital director's statement was the truth (he referred to the ward by its current name), but not the whole truth. He himself was responsible for the name-change of the ward.  And later: the year after the quiet year after the big year, the numbers were big again.

Amusingly, in the case of Lucia de Berk, the argument put by the prosecution was precisely that respiratory arrest
was normal, while cardiac arrest was supposed to be unusual! In one of the key events, the crucial question for deciding whether a baby had died of poisoning or naturally, was whether heart failure took place before or after lung failure. The argument being: if someone is terminally ill, then the natural course of events is that the body becomes exhausted, the lungs fail, consequently there is shortage of oxygen, then the heart fails. On the other hand, ``unexpected'' heart failure (after which the lungs naturally fail, too) could indicate poisoning.

It turned out that the temporal sequence of events was not remembered in the same way by different observers (doctors, nurses) and moreover that different registration systems appeared to give a different answer. Only after an extremely carefully and thorough investigation taking everything into account, could it be concluded that quite definitely, respiratory arrest came first, followed by heart failure. 

I don't suppose that how terminally ill people die in the Netherlands is terribly different from how they do it in the UK. And I apologise for appearing to offer anecdotal medical evidence when I am a statistician. However this point is crucial: medical diagnosis is not an exact science! Many things happen in rapid succession. What one takes as the ``cause of death'' or the ``cause of the emergency'' is \emph{fuzzy}. Impressions of skilled doctors might easily be different from the information obtained from fairly reliable medical monitoring systems. The output from different monitors can get confused and mis-interpreted. Memories are unreliable. Memories change, certainly when people start to believe that there is a killer around. 

These points are actually very relevant to any interpretation of the statistics in the case of Ben Geen. Which events are registered as events of which category is to a large degree subjective, and variable. 

I would also like to mention an ``incident'' pertaining to the present case. At my request, an F.O.I.\ request was also put
to Horton General, sometime later than all the others.  The data which this hospital was able to prepare pertained only to a very short, recent, period, and with almost all categories merged. All numbers of events in any month smaller than 5 were simply reported as ``< 5''. No other hospital took this weird precaution, allegedly taken for reasons of confidentiality. Is this an attempt to suppress embarrasing evidence?
\newpage
\section{Appendix 4: Overview of our data sources}
\begin{center}
\noindent \includegraphics[width = 5.5 in]{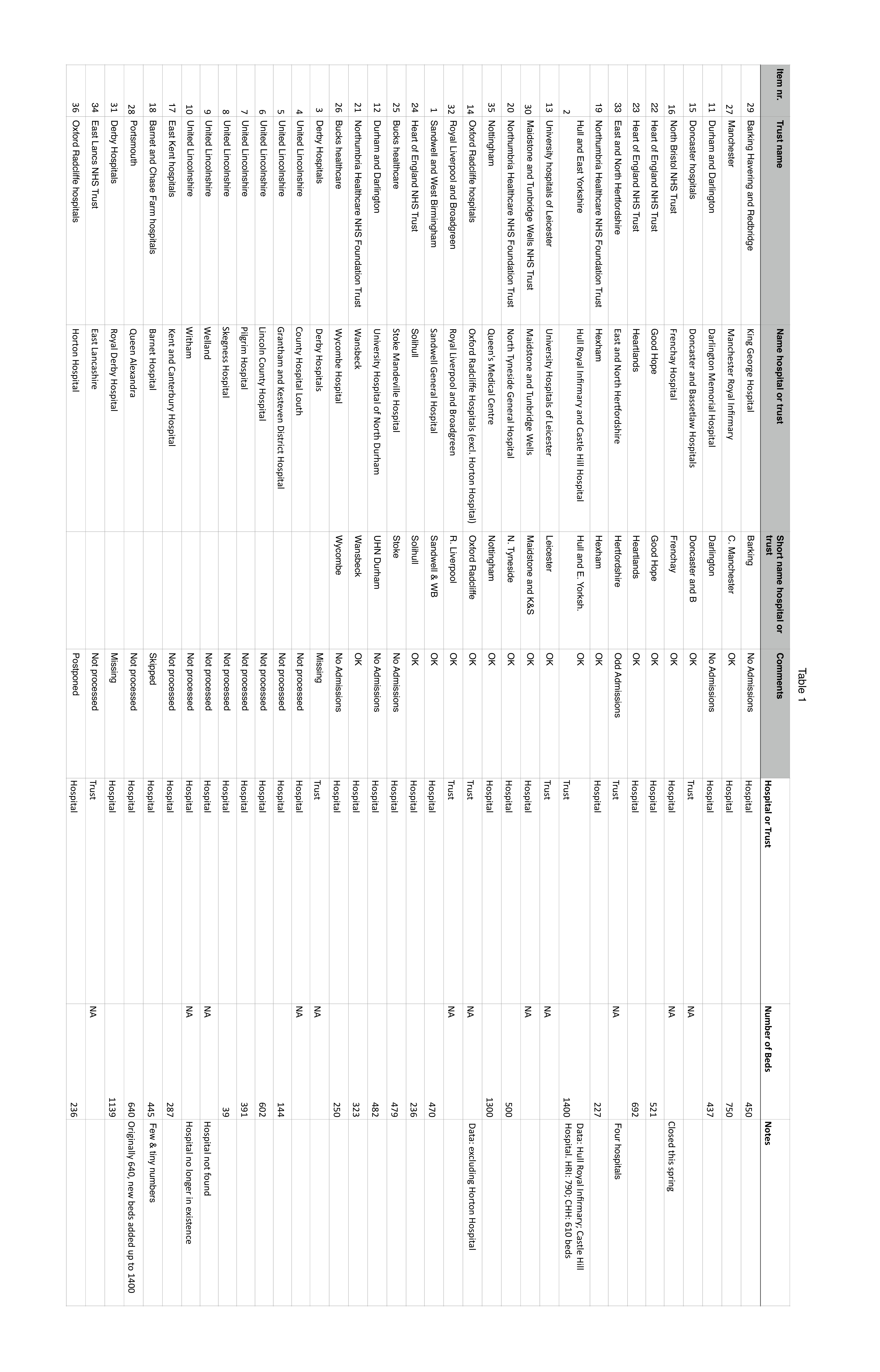}
\end{center}

\end{document}